# High $T_C$ ferromagnetism in diluted magnetic semiconducting GaN:Mn Films

Hidenobu Hori[1], Saki Sonoda[2], Takahiko Sasaki[1], Yoshiyuki Yamamoto[1], Saburo Shimizu[2], Ken-ichi Suga[3] and Koichi Kindo[3]

[1] *School of Materials Science, Japan Advanced Institute of Science and Technology (JAIST), 1-1 Asahidai Tatsunokuchi, Ishikawa 923-1292, Japan*
[2] *ULVAC, Inc., 2500 Hagisono, Chigasaki, Kanagawa 253-8543, Japan*
[3] *Research Center for Materials at Extreme Conditions (KYOKUGEN), 1-3 Machikaneyama, Osaka University, Toyonaka, Osaka 560-8531, Japan*

**Abstract**

Wurtzite GaN:Mn films on sapphire substrates were successfully grown by use of the molecular beam epitaxy (MBE) system. The film has an extremely high Curie temperature of around 940 K, although the Mn concentration is only about 3 ~ 5 %. Magnetization measurements were carried out in magnetic fields parallel to the film surface up to 7 T. The magnetization process shows the coexistence of ferromagnetic and paramagnetic contributions at low temperatures, while the typical ferromagnetic magnetization process is mainly observed at high temperatures because of the decrease of the paramagnetic contributions. The observed transport characteristics show a close relation between the magnetism and the impurity conduction. The double exchange mechanism of the Mn-impurity band is one of the possible models for the high-$T_C$ ferromagnetism in GaN:Mn.

**Corresponding to:** Prof. H. Hori
Address: School of Materials Science, Japan Advanced Institute of Science and Technology (JAIST), 1-1 Asahidai, Tatsunokuchi, Ishikawa 923-1292, Japan
E-mail: h-hori@jaist.ac.jp
TEL: +81-761-51-1550
FAX: +81-761-51-1535

# 1. Introduction

Research on the diluted magnetic semiconductors has attracted much interest since the first fabrication of Mn doped InAs, (InAs:Mn) [1]. The result has led to intensive investigations on the III-V based Diluted Magnetic Semiconductors, DMS [2-4]. From application point of view, some ferromagnetic semiconductors are expected to be candidates for the spin polarizer [5], which is a ferromagnetic material for the spin injection to a non-magnetic semiconductor. However, the Curie temperature $T_C$ of the GaAs-based DMS is presently as low as 110 K. This fact means that the function is not available at room temperature. In the mean while, recent development of the growth techniques of the wurtzite type Gallium-nitrides has resulted in the successful fabrication of the GaN-based optical and electrical devices [6, 7]. Moreover, there are some theoretical predictions of the possibility of GaN-based DMS having Curie temperature exceeding room temperature [8, 9]. There are some experimental studies [10-14] on magnetic characteristics of the GaN-based DMS and none of these studies was able to show the sample having the $T_C$ exceeding room temperature. Especially, Ref. [14] observed anomalous "offset magnetization" at room temperature related with the ferromagnetism, but they did not conclude the ferromagnetism. However, our previous work [15] reported on ferromagnetic characteristics of the GaN:Mn film by an observation of magnetization process demonstrated clear hysteresis loop. In addition to the hysteresis loop arising from the ferromagnetic domain structure, clear step like magnetization around zero field also demonstrated the ferromagnetism. The Curie temperature of much higher than 400 K was estimated from the temperature dependent saturation moment.

The first purpose of this study is to discuss the magnetic characteristics of the DMS films. The second one is the discussion of the coexistence of the ferromagnetic and the paramagnetic parts as well as the origin of the high Curie temperature $T_C$ in the GaN:Mn. The third one is the discussion for a tentative but most possible model based on the experimental results to the high $T_C$ ferromagnetism on GaN:Mn.

## 2. Experimental procedures and results

### 2.1 Sample preparation

In this section, we report a successful growth of wurtzite GaN:Mn film exhibiting ferromagnetic behaviour having Curie temperature above RT. The films were grown by Molecular Beam Epitaxy using ammonia nitrogen source ($NH_3$-MBE) system (ULVAC, MBC-100) equipped with a Reflection High Energy Electron Diffraction (RHEED) apparatus. Solid-source effusion cells were used for Ga and Mn sources. GaN:Mn films of thickness ranging from 1300 to 6000 Å were grown at temperatures between 850 and 1020 K and at various Ga/Mn flux ratios on wurtzite GaN layers on sapphire(0001) substrates [15]. Observed RHEED pattern revealed that single crystal wurtzite GaN:Mn films are obtained at a higher growth temperature and at a higher Ga/Mn flux ratio. Also zinc-blende crystals were included in the films grown at a lower growth temperature and at a lower Ga/Mn flux ratio. The growth condition dependence of the crystal structures of the grown films will be presented in detail in Ref. [16]. After the growth of the GaN:Mn film, a pure GaN layer with 200 Å–thickness was grown as a cap layer to prevent the oxidation of the film. The quality of GaN:Mn films was examined by RHEED and X-Ray Diffraction (XRD) before measuring magnetic and electrical properties. Table I gives growth conditions and the RHEED pattern observed during the growth of the GaN:Mn film and the GaN buffer layer. XRD measurements revealed that the film had the wurtzite structure without phase separation [16]. The wurtzite type crystal structure of the GaN:Mn film was confirmed by Coaxial Impact Collision Ion

Scattering Spectroscopy (CAICISS). Moreover, CAICISS measurement clearly denied any possibility of segregation on the surface of the GaN:Mn film within the resolution limit. The results of the detailed CAICISS measurement were discussed in Ref. [17]. These were consistent with the observed RHEED patterns of the GaN:Mn film. From the depth profile measured by a Secondary Ion Mass Spectroscopy (SIMS) and spectra measured by X-ray Absorption Fine Structure (XAFS), it was confirmed that the Mn concentration in the GaN:Mn film was uniform within the resolution limit of 30 nm and Mn atom was incorporated at the Ga site [16, 18]. The sample of GaN film with 3000 Å thickness on a GaN buffer layer was prepared for a comparative study of the magnetic properties. It was used to subtract the stray magnetization of the GaN layer with the sapphire substrate from the observed magnetization data of GaN:Mn film. The Mn-concentration in the samples used in this work was about 3 ~ 5%. The concentration was estimated by SIMS study (Model IMS-5F, CAMECA Co. Ltd.) equipped at JAIST.

**2.2 Magnetization measurement**

Magnetization measurements of the GaN:Mn films were carried out by a Superconducting Quantum Interference Device (SQUID) magnetometer (MPMS-XL, Quantum Design Co. Ltd.) in a temperature range from 1.8 to 750 K. Magnetic fields up to 7 T were applied parallel to the film plane, because the spins are in-plane. In order to determine the $T_C$ that appeared to be higher than 400 K, an oven having precise temperature control was employed in the SQUID magnetometer. The background contribution to the magnetization is originated from the GaN buffer layer, cap layer and from the diamagnetism of the sapphire substrates. These were experimentally eliminated from the original data of the sample.

Figure 1 shows the temperature dependence of the magnetization ($M$-$T$ curve) in 0.1

and 7 T. The magnetic field of 0.1 T, which makes negligible change in paramagnetism, is just above the field corresponding to the top of the hysteresis loop measured at 1.8 K. Therefore, $M$-$T$ curve in Fig.1 is considered to be the temperature dependence of the spontaneous polarization, $M_{Sp}$ in the ferromagnetic state. The $M$-$T$ curve is observed up to 750 K to precisely estimate the value of $T_C$.

The field dependence of the magnetization, *i.e.* the magnetization process ($M$-$H$ curve) was measured up to 400 K. Typical examples of the data are shown in Fig. 2. The hysteresis loop is visible in the $M$-$H$ curve at these temperatures. The value of coercive field at 300 K shown in the inset of Fig. 2 is about 6 mT. A steep increase in the $M$-$T$ curve at 4.2 K looks like "coexistence" of paramagnetic and ferromagnetic components. The feature of the coexistence is also seen in the $M$-$H$ curve, which looks paramagnetic like magnetization process on the ferromagnetic magnetization. However, the steep increase in $M$-$T$ curve does not mean a simple "coexistence", because the saturation magnetization, $M_{Top}$ at the top of the hysteresis loop also has the steep increase, as indicated by open squares in Fig. 1. The steep increase of the saturation magnetization was directly confirmed by the observation of anomalous increase of the remanent moment in the same low temperature range. The magnetization and susceptibility data will have much information, but the careful analysis is required for the quantitative discussion, because the ferromagnetism generated by doping Mn ions is not a simple coexistence of the paramagnetic and ferromagnetic components. These phenomena prevent the clear decomposition between paramagnetic and ferromagnetic components. A careful analysis on the detailed temperature dependence of the magnetization process will be given by Suga *et al*. in near future [19]. The results suggest the temperature dependence of the ferromagnetic saturation moment or change in the exchange interaction. Such "coexistence" characteristics have been commonly observed in the experiment of other similar DMS materials [20]. We believe that these characteristics are intrinsic problem of DMS materials. To the best of our knowledge, an elaborate and systematic discussion has not appeared yet.

The magnetization curve in 7 T is close to the curve of the "ferromagnetic spontaneous magnetization" in 0.1 T at room temperature. The approach mainly arises from the temperature dependence of the paramagnetic component. Analysis of inverse susceptibility plot is not appropriate in case of the high $T_C$ ferromagnet like GaN:Mn, because the high temperature expansion of the plot is not applicable to the magnetic coexistence system having temperature dependent saturation moment. From these results, it can be stated that the GaN:Mn film shows ferromagnetism and the Curie temperature is higher than 750 K although the coexistence with "paramagnetism" is observed. The value of $T_C$ of the present sample is quite high in comparison with other ferromagnetic DMS materials reported until now. The approximate value of $T_C$ was estimated to be 940 K by use of a usual theoretical curve given by the molecular field approximation [21].

**2.3 Discussion on segregation problem**

It may be inferred that the segregation of some ferromagnetic compounds may be an origin of the high $T_C$ ferromagnetism of this sample. The possibility, however, is clearly denied by the following reasons: 1) The segregated small ferromagnets should show characteristics of the superparamagnetism. The observed ferromagnetic *M-H* curve in this study is completely different from the superparamagnetism. Especially, it is a typical ferromagnetic curve above room temperature. 2) Uniform distribution of Mn in the sample was confirmed within the resolution limit of the SIMS and Rutherford Back Scattering, RBS (Model 1700H, Nisshin-High Voltage Co. Ltd.) equipped at JAIST.

The resolution limits of these equipments are approximately 30 and 20 nm, respectively. The result of CAICISS experiment also clearly shows uniform surface without segregation. 3) Strong fluorescence of GaN observed at 3.3 eV (in ultra-violet region) is completely disappears in GaN:Mn. If the segregation is generated by the 3% Mn ions on the sample, the optical effect of the segregation should be appeared as the screening of the fluorescence with same wavelength. However, such fluorescence is not observed. Instead of the screened spectra, quite weak one is observed at different wavelengths around 2 eV (orange color region). The result strongly suggests an intrinsic change in the electronic state made by the Mn-doping, because such a change in electronic state cannot be made by the coexistence of GaN and segregated materials. 4) The change in carrier type and large variation in carrier density from *n*-type GaN to *p*-type GaN:Mn support the intrinsic change of the electronic state made by the Mn-doping. 5) The Curie temperature of the present sample is much higher than the highest $T_C$ of 748 K [22] in Mn-Ga alloying system and $T_C$ of 720 K [23] of $Mn_4N$. These are possible materials made from Ga, N and Mn. The α-Mn metal and the ionic compound MnO are also possible compounds but they are antiferromagnetic. These experimental results strongly support that the GaN:Mn grown by the MBE in this work is new intrinsic ferromagnet of DMS with high $T_C$.

## 2.4 Transport measurements

Because GaN:Mn is one of the DMS materials, the origin of the ferromagnetism is considered to be closely related with the properties of the carrier. In this section, it is discussed that the carrier transport is dominated by hopping conduction through the impurity band of Mn. The temperature range of the trapping effect of the carriers is consistent with that of the anomalous increase of the magnetization and the results make us infer that there is a close relation between hopping carrier and magnetization.

For the composite of GaN-buffer layer/ GaN:Mn film/ GaN-cap layer, the transport measurements are already reported [24]. In order to obtain more precise transport characteristics of GaN:Mn film, samples having no-cap layer, relatively thick GaN:Mn film and buffer-layer with high resistance were specially prepared. The ferromagnetism at room temperature of these samples is confirmed before the transport measurement. Sn metal was used to make the ohmic contact between the sample and lead wire. The magnetoresistance measured by four terminal method and Hall resistance are shown in Fig. 3(A) and (B). The net carrier concentration was estimated from the data of Hall resistance. As is seen in the raw data, symmetric field dependence for zero-field shown in the inset of Fig. 3(B) is unnatural for the Hall resistance data. For this result, it can be considered for this result that the magnetoresistance part straying into the Hall resistance data because of unavoidable miss alignment of the electrodes. The stray magnetoresistance is eliminated by comparing two Hall resistance data which were taken by use of two opposite magnetic fields, keeping other circuit configuration unchanged.

Figure 4(A) shows the temperature dependences of the resistance and of carrier density. The carrier density is estimated from the Hall resistance. Both Hall resistance and magnetoresistance data were observed in fields higher than 7 T. The high magnetic field is applied in order to avoid the domain and ferromagnetic diamagnetic effect in anomalous field dependence region. Like Ref. [8], the sign of majority carrier in this work is found to be positive and it is a hole-dominant *p*-type conductor.

It is worth mentioning that coincidence between the anomalous increase of spontaneous magnetization and the decrease of carrier density $n$ approximately below 10 K. The result suggests a close relation between magnetism and carrier. The plots of the resistivity $\rho$ and carrier density $n$ and their logarithmic plots are shown in Fig. 4. The temperature dependence of the plot indicates trapping characteristics in liquid He temperature and shows the existence of some trapping states in GaN:Mn. The temperature dependence of the charge density is relatively gentle in the high temperature region as shown in Fig. 4(A). Because the intrinsic gap energy of 3.4 eV in GaN is quite large, it can be considered that the high temperature region is so-called exhaustion range in the impurity conduction [25].

The trapping energy in the lowest temperature range is estimated to be about 0.04 meV. This result means existence of some quite shallow trapping potential. The decrease of carrier density affects on the conductivity or resistance. In the case of *p-d* scattering model on the delocalized carrier system, such a decrease of carrier density with decreasing temperature cannot be considered although the resistance increase with decreasing temperature (Kondo effect). The following idea of weak localization mechanism is applicable for the shallow localized level: the randomness of the distribution of impurity ions makes sharp localized level at low temperatures but thermal motion at higher temperatures makes broadening and hybridization of the localized level and the impurity band is formed. Experimentally, as shown in the inset of Fig. 3(A), the conductivity proportional to log *T* below 20 K supports the weak localization model.

Generally, the ferromagnetism in high $T_C$ 3*d*-conductors is closely related with the strong magnetic interaction between carriers in narrow 3*d*-band originated from localized magnetic cores. In case of GaN:Mn, the most possible origin of the shallow level is the acceptor of Mn in GaN and the hopping carriers on Mn ions in higher temperature region form an impurity band. The spin dependent hopping band of Mn impurities near the Fermi level in GaN:Mn is considered to be the origin of the ferromagnetism. The quite shallow energy gap of 0.04 meV is easy to understand by the weak localization model. The hopping conduction mechanism on Mn impurities makes us infer the relatively large magnetic scattering. In fact, as is seen in Fig 3(A), the large change in magnetoresistance is found in low temperature range.

**3. Discussion and Conclusion.**

The magnetic experiments of GaN:Mn conclude a coexistence of paramagnetic and ferromagnetic components. The highest Curie temperature of the ferromagnetism observed in this work is about 940 K and a clear hysteresis loop is observed at temperature as high as 400K. Because the coexistence of ferromagnetic and paramagnetic parts is generally observed in DMS materials [13], the origin is quite general and intrinsic problem. The most characteristic property in DMS materials is the random distribution of magnetic impurities. The random distribution of magnetic impurities includes some lone impurities although other magnetic impurities have ferromagnetic coupling. Such lone magnetic impurities are considered to be the origin of the paramagnetism. The important and interesting problem is why the magnetic impurities make such a high $T_C$ ferromagnetism in DMS materials. The high $T_C$ ferromagnetism is related with the strong exchange coupling among the magnetic impurities in the impurity band.

The localized properties of the carriers are remarkably appeared at low temperatures. In fact, as discussed in the previous section, anomalous increase of the spontaneous magnetization in the lowest temperature range below 10 K is consistent with the decreasing region of the carrier density in the similar temperature range. This means the trapping of the electrons on Mn impurities makes increase of magnetization.

Although the origin of high $T_C$ ferromagnetism has not yet been confirmed well, a tentative model is proposed here to make consistent explanation of the experimental results. The model is based on the band theory of DMS given by Sato and Katayama [9]. According to the theory, the narrow 3$d$-band on Mn impurities in GaN is generated near the Fermi level. The impurity band is originated from the anti-bonding orbits of $t_{2g}$-level in Mn impurities. It should be emphasized that the anti-bonding orbit makes spin-triplet state, that is ferromagnetic spin transfer between Mn ions and this is the origin of the double exchange mechanism. The theory predicts slight energy gap between Fermi energy and the anti-bonding $t_{2g}$-band for the ideal GaN crystal. But it is noticed that the actual GaN has tendency to be $n$-type semiconductor. In fact, GaN produced by NH$_3$-MBE method has electron density of $10^{26}$ m$^{-3}$. The electron density in the $n$-type GaN is about 10 times larger than the hole density of $p$-type GaN:Mn. Therefore, it is easy to consider that the excess electrons can make the Fermi energy pull up to the $t_{2g}$-band and the magnetic scattering can produce the ferromagnetism.

A tentative model is introduced here to explain the high $T_C$ ferromagnetism as follows: the spin splitting or shift between up- and down- spin bands of hopping carriers is generated by double exchange interaction. The double exchange mechanism is well known and described in standard textbook of magnetism. The ferromagnetic parallel spin hopping is originated from anti-boning orbital in Mn impurities. The characteristics of the strong exchange interaction in GaN:Mn arises from the mechanism of the double exchange interaction, because the double exchange interaction is so-called direct exchange interaction between Mn spins. This means the interaction is lower rank perturbation in comparing with the indirect exchange interaction such as $p$-$d$ (or $s$-$d$) interaction discussed in Ref. [8]. Namely, the direct exchange interaction does not pass through the conduction electron.

This tentative model also explains the field and temperature dependence of the magnetization in the lowest temperature region. Such a localization of the hopping electron prevents the hopping and reduces the ferromagnetic double exchange interaction. Therefore, the ferromagnetic band splitting is also destructed. The localization also makes increase of the magnetic moment. This effect is easily understood, if the hopping electrons are trapped and back to the divalent Mn states. The trapping process makes increase of the localized moment because the spin polarization per electron in conduction band is usually smaller than localized one. This is the reason why the exchange interaction between Mn ions is reduced at low temperatures, while the localized moment is larger than ferromagnetic phase. Therefore, this model consistently explains the coincidence of the temperature region between decrease of carrier density and increase of magnetization.

The tentative model requires the concentration dependence of $T_C$ and magnetization. In fact, large concentration dependence of the saturation moment has been observed. However, the experimental error of the magnetization curve prevents the exact determination of $T_C$. Therefore, the relation between concentration of Mn and $T_C$ has not been obtained, so far. The experimental error is mainly due to the low sensitivity in the measurement by oven system in SQUID magnetometer. In order to make clear the origin of High $T_C$ ferromagnetism, more detailed and high sensitive experimental method should be developed for the various Mn-concentration samples. Such a problem is opened for future investigation.

**Acknowledgements**

This work is partly supported by the Ministry of Education, Culture, Sports, Science and Technology, Government of Japan (MONBUKAGAKUSHO).

**Table caption**

Table 1. Growth conditions and RHEED pattern of the GaN:Mn film and GaN buffer layer.

|  | GaN | GaN:Mn |
|---|---|---|
| Cell temperature (K) |  |  |
| Ga | 1170 | 1120 |
| Mn | - | 850 |
| $NH_3$ flow ($10^{-3}$ Pa·m$^3$/s) | 8.4 | 8.4 |
| Substrate temperature (K) | 990 | 990 |
| Thickness (Å) | 2000 | 3600 |
| RHEED pattern | 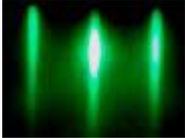 | 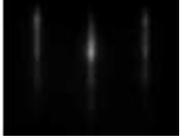 |

**Figure captions**

Fig. 1. Temperature dependence of the magnetization (*M-T*) at 0.1 (closed circles) and 7 T (open circles) in the temperature range between 1.8 and 300 K. Open squares represent the magnetization at the magnetic field of the top of the hysteresis curve measured at each temperature. The inset shows the temperature dependence of the magnetization up to 750 K and the solid line represents the calculation curve by use of the molecular field approximation. The ferromagnetic transition temperature $T_C$ is estimated to be 940 K.

Fig. 2. Magnetic field dependence of the magnetization (*M-H*) at 1.8 and 300 K. The broken line represents the magnetic field of the top of the hysteresis loops. The inset shows the magnetization processes up to 7 T.

Fig. 3. (A) The magnetoresistance data at various temperatures between 1.8 and 200 K. The magnetic field was applied parallel to the film plane and swept between –10 and +10 T. The inset shows the logarithmic temperature dependence of the conductivity. The broken line is the fitting curve representing log *T* dependence. (B) The Hall resistance data at various temperatures between 1.8 and 200 K. The magnetoresistance contribution to the Hall resistance was subtracted by use of two raw Hall resistance data at opposite magnetic fields. The inset shows corresponding raw data.

Fig. 4. (A) Temperature dependence of the resistivity and the carrier density estimated from the Hall resistivity measurement. (B) The logarithm of the resistivity and the carrier density as a function of the inverse of temperature.

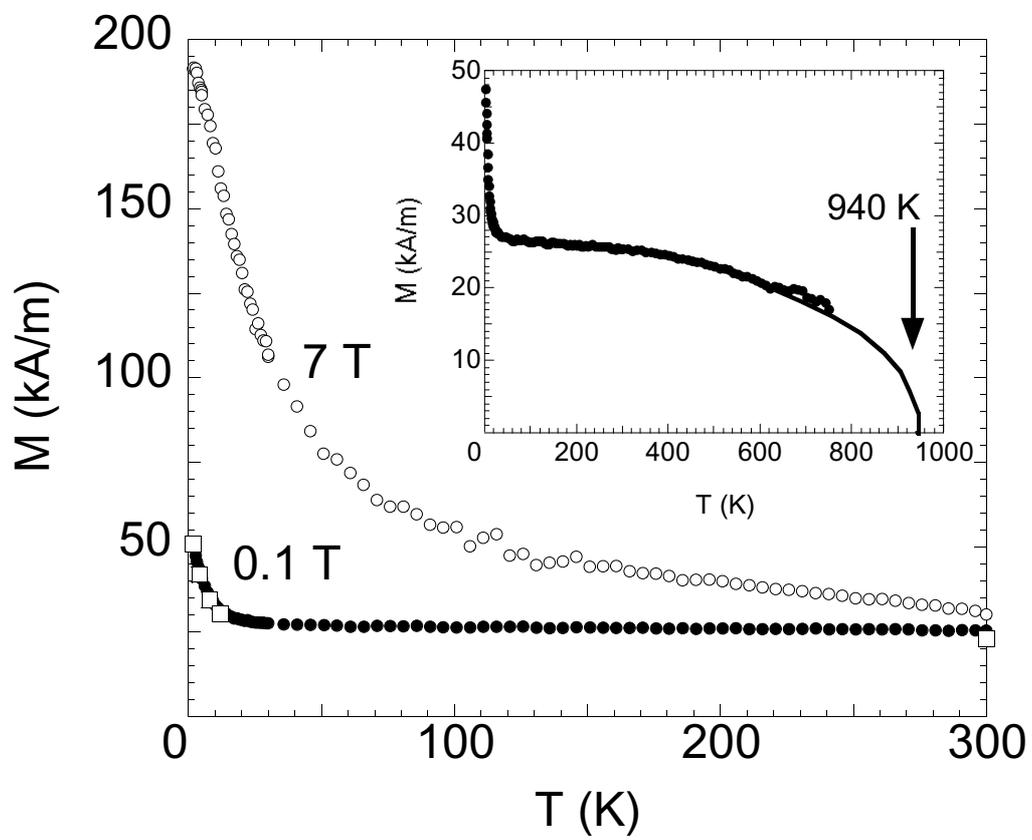

Fig.1 H. Hori et al.

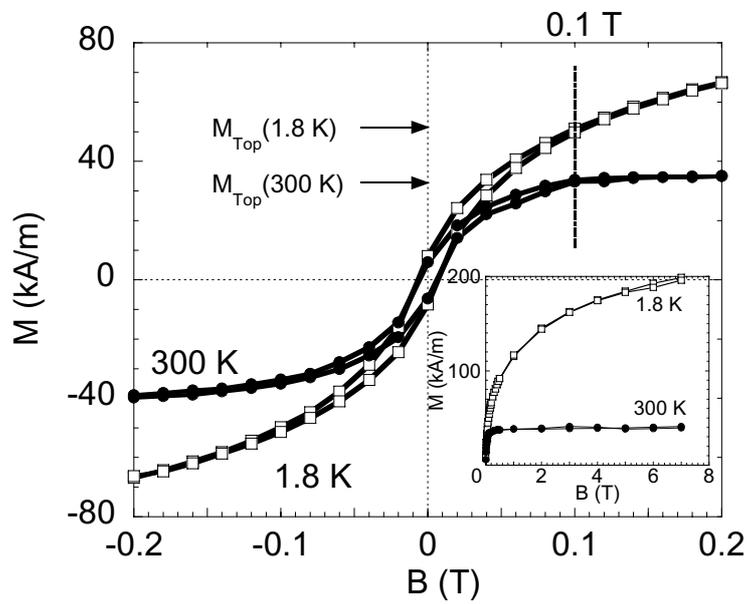

Fig. 2 H. Hori et al.

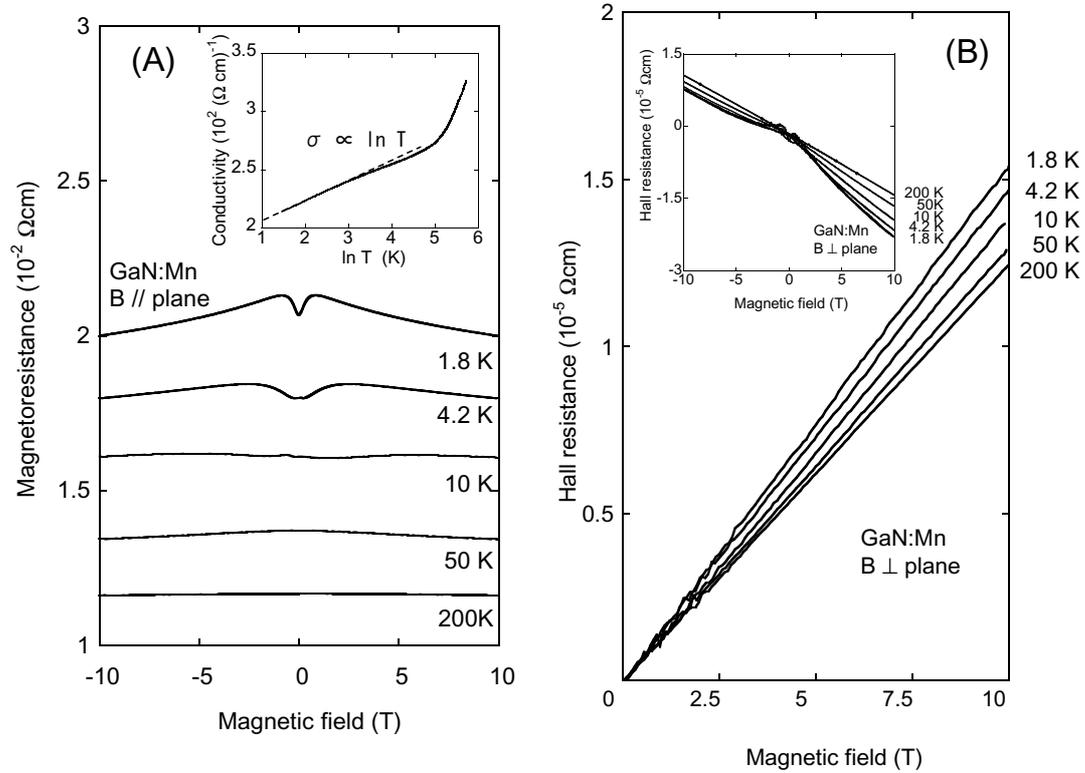

Fig. 3 H. Hori et al.

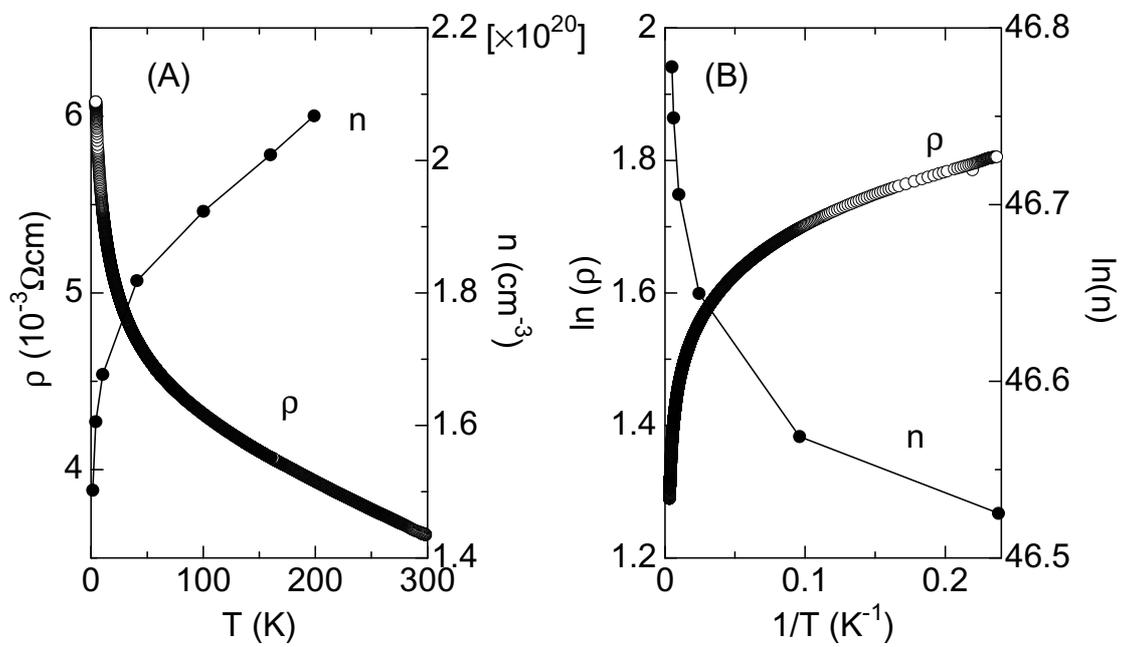

Fig. 4 H. Hori et al.